\documentclass[fleqn,usenatbib]{mnras}
\usepackage{newtxtext,newtxmath}
\usepackage[T1]{fontenc}
\usepackage{ae,aecompl}

\usepackage{graphicx}	% Including figure files
\usepackage{amsmath}	% Advanced maths commands
\usepackage{amsfonts}
\usepackage{color,soul}
\usepackage[table]{xcolor}
\usepackage{tabu}
\usepackage{tikz}
\usetikzlibrary{calc}

\title[Classification and evolution of cluster galaxies]{Classification and evolution of galaxies according to  the dynamical state of host clusters and galaxy luminosities}

\author[D. F. Morell et al.]{
D. F. Morell,$^{1,2}$\thanks{E-mail: dailerfm@gmail.com}
A. L. B. Ribeiro,$^{2}$
R. R. de Carvalho,$^{3}$
S. B. Rembold,$^{4}$\and
~P. A. A. Lopes,$^{5}$
A. P. Costa$^{2}$\\
$^{1}$Observat\'orio Nacional -- MCTIC, Rio de Janeiro, RJ, 20921-400, Brazil\\
$^{2}$Universidade Estadual de Santa Cruz, Laborat\'orio de Astrof\'isica Te\'orica e Observacional, Ilh\'eus, BA, 45650-000, Brazil\\
$^{3}$NAT - Universidade Cruzeiro do Sul / Universidade Cidade de S\~ao Paulo, 01506-000 Brazil\\
$^{4}$Universidade Federal de Santa Maria, Santa Maria, RS, 97105-900, Brazil\\
$^{5}$Observat\'orio do Valongo, Universidade Federal do Rio de Janeiro, Rio de Janeiro, RJ, 20080-090, Brazil
}

\date{Accepted XXX. Received YYY; in original form ZZZ}

\pubyear{2019}

\begin{document}
\label{firstpage}
\pagerange{\pageref{firstpage}--\pageref{lastpage}}
\maketitle

\begin{abstract}
We analyze the dependence of galaxy evolution on cluster dynamical state and galaxy luminosities for a sample of 146 galaxy clusters from the Yang SDSS catalog. Clusters were split according to their velocity distribution in Gaussians (G) and Non-Gaussians (NG), and further divided by luminosity regime. We performed a classification in the Age-SSFR plane providing three classes: star-forming (SF), passive (PAS), and intermediate (GV -- green valley). We show that galaxies evolve in the same way in G and NG systems, but also suggest that their formation histories leads to different mixtures of galactic types and infall patterns. Separating the GV into star-forming and passive components, we find more bright galaxies in the passive mode of NG than in G systems. We also find more intermediate faint galaxies in the star-forming component of NG than in G systems. Our results suggest the GV as the stage where the transition from types Sab and Scd to S0 must be taking place, but the conversion between morphological types is independent of the dynamical stage of the clusters. Analyzing the velocity dispersion profiles, we find that objects recently infalling in clusters have a different composition between G and NG systems. While all galaxy types infall onto G systems, Sab and Scd dominate the infall onto NG systems. Finally, we find that faint Scd in the outskirts of NG systems present higher asymmetries relative to the mean asymmetry of field galaxies, suggesting environmental effects acting on these objects.

\end{abstract}


\defcitealias{rhee2017phase}{R17}
\defcitealias{mahajan2011velocity}{M11}
\defcitealias{decarvalho}{dC17}

\newcommand{\tm}{\textrm{}}
\newcommand\rbd[1]{\color{red}{\textbf{#1}}}
\newcommand{\ccl}{\cellcolor{blue!25}}
%\newcommand{\ccl1}{\cellcolor{red}}

%%%%%%%%%%%%%%%%%%%%%%%%%%%%%%%%%%%%%%%%%%%%%%%%%%

%%%%%%%%%%%%%%%%% BODY OF PAPER %%%%%%%%%%%%%%%%%%

\section{Introduction}

%Galaxy clusters are massive objects made of few hundreds of galaxies, virialised
%in their core where massive and old galaxies usually resides. 
%Galaxies infalling towards the center of a cluster crosses it several times before settling in the central regions,
%with typical crossing time of  $\sim$1.7 Gyr for rich clusters \citep[e.g.]{boselli2006environmental}. Hence,

Galaxy clusters are valuable laboratories to understand galaxy evolution in connection with the environment.
In the hierarchical scenario, massive structures should be forming at the present epoch, as a result of mergers and accretion of less massive objects. Thus, clusters constitute complex systems in
a wide range of dynamical and evolutionary states, usually reduced to two categories: virialized (or relaxed) and
non-virialized (or non-relaxed) systems.
Virialized clusters are expected to have nearly spherical shape and gaussian velocity distribution \citep[e.g.][]{yahil1977velocity, faltenbacher2006velocity}, while non-virialized systems can show elongated shapes \citep{plionis2004large}, non-gaussian velocities \citep[e.g.][]{hou2009statistical,2013MNRAS.434..784R}, and the presence of substructures \citep[e.g.][]{roberts18,lopes18}.
Some works suggest that the way galaxies evolve is dependent on the dynamic state of their host clusters
\citep[e.g.][hereafter dC17]{hou2009statistical,2012MNRAS.421.3594H,2013MNRAS.434..784R,decarvalho}, which underlines the importance of correctly determining the dynamic state of the clusters.

Different methods for assessing the relaxation degree of galaxy systems are found in the literature. 
For instance, detection of optical substructures \citep[e.g.][]{2012MNRAS.421.3594H,wen13,cohen15,lopes18,soares19}, deviation of scale relations \citep[e.g.][]{yang10}, merger signatures on images \citep[e.g.][]{plionis03,allen08,million09}, X-ray peak and BCG offset \citep[e.g.][]{rossetti16,roberts18,lopes18,foex19}, X-ray photon asymmetry and/or centroid shift \citep[e.g.][]{ge18,roberts18,bartalucci19}. 
In addition to these methods, the shape of the radial velocity distribution of cluster galaxies has been proved to be robust for separating relaxed from unrelaxed clusters, where ``relaxed" means a system having member-galaxy velocity distribution roughly following the Maxwell-Boltzmann function \citep[e.g.][dC17]{2013MNRAS.434..784R}. Recently, \citet{roberts2018connecting} demonstrated that clusters with velocity distributions well fitted by a Gaussian have X-ray morphologies which are usually symmetric, whereas clusters with non-Gaussian velocity profiles show X-ray morphologies with significant asymmetries, suggesting the use of velocity distributions as a reliable way to determine cluster dynamical state.

When assessing Gaussianity in the radial velocity distribution, Gaussian and Non-Gaussian systems exhibit significant differences \textit{wrt} some of their galaxy properties \citep{ribeiro10b,2013MNRAS.434..784R,ribeiro13}. For example, member galaxy luminosities are related to the dynamical state \citep[e.g.][]{martinez12,wen13},
with non-Gaussian systems presenting a fainter characteristic absolute magnitude \citep{martinez12}.
 Also, non-Gaussian groups have larger velocity dispersions and, as a consequence, higher dynamical masses \citep{ribeiro11,old2018galaxy}. In addition, less relaxed clusters show an increased star formation rate \citep{cohen15,2017MNRAS.467.3268R}, where faint galaxies are mainly find in the cluster outskirts, and being more metal rich, which can be interpreted as an evidence of preprocessing \citepalias{decarvalho}. Regarding the specific star formation rate (SSFR), \citet{nascimento19} find that the Gaussian passive population presents velocity segregation with luminosity. In addition, \citet{soares19} find lower mean stellar ages in unrelaxed clusters relative to relaxed clusters, and explain the differences between them as mainly driven by low-mass systems in the clusters outskirts, where substructures lies preferentially \citep[e.g.][]{2012MNRAS.421.3594H}. These results suggest a connection between the dynamical state of clusters and the processes driving galaxy evolution, a link yet to be fully understood.

The cluster dynamical state is related to the orbits of member galaxies. As a galaxy moves through the cluster, its orbit is eventually modified, from highly radial trajectories in the cluster outskirts to circular orbits in central regions inside the virial radius. Along with galaxy dynamical history, physical properties also changes influenced by the cluster environment, and so can be linked to the orbital type. For example, different Hubble types are associated with different orbits, with early-type galaxies having more isotropic orbits than late-type galaxies \citep[e.g.][]{tammann72,sodre89,adami98,biviano2002eso,aguerri07,cava17,mamon2019structural}. 
The velocity dispersion profile (VDP) shape is also related to some properties such as the efficiency of merger activity and/or substructure \citep[e.g.][]{menci96,2012MNRAS.421.3594H,pimbblet14,bilton18}, presence of different spectral classes galaxies \citep{rood72} and cluster dynamical state \citep[e.g.][]{hou2009statistical,costa17,nascimento19}.

This paper is the fourth of a series in which we investigate the discrimination
between Gaussian (G) and Non-Gaussian (NG) clusters, based on the velocity distribution of the member galaxies. First, \citet{decarvalho} report evidence that faint
galaxies in the outer regions of NG groups are infalling for
the first time into the systems. Then, \citet{costa17} 
find significant differences between
the velocity dispersion profiles (VDP) of G and NG system, and also 
regarding bright and faint samples of galaxies.
 Finally, the study of \citet{de2019mass} on the Star Formation History (SFH) of cluster galaxies reveals 
that the SFH in faint spirals of NG groups is significantly different from
their counterpart in the G groups, leading the authors to conclude that there is a higher infall rate of gas rich systems in NG groups.

%This paper is the fourth of a series in which we investigate the discrimination
%between Gaussian (G) and Non-Gaussian (NG) clusters, based on the velocity distribution of the member galaxies. First, \citet{decarvalho} find that 76\%
%of the Yang's groups \citep{yang07} with masses above $10^{14}\;{\rm M_\odot}$ have Gaussian velocity distributions. They also find evidence that faint
%galaxies in the outer regions of NG groups are infalling for
%the first time into the systems. Then, \citet{costa17} studied
%the velocity dispersion profiles (VDP) of G and NG systems and find that the stacked VDP for G groups exhibit a central peak followed by a monotonically decreasing behavior indicating predominantly radial orbits, while the Bright stacked VDPs show lower velocity dispersions than the Faint
%stacked VDPs, in all radii. As far as NG systems are concerned, they display a distinct feature with a depression in the central region and also a likely higher infall rate associated with galaxies
%in the Faint stacked VDP. Finally, the study of \citet{de2019mass} on the Star Formation History (SFH) of ellipticals and spirals reveals 
%that the star formation history in faint spirals of NG groups is significantly different from
%their counterpart in the G groups. The assembled mass for faint spirals varies from
%59\% at 12.7 Gyr to 75\% at 8.0 Gyr, while in G systems this variation is from 82\% to
%91\%. This finding may also be interpreted as a higher infall rate of gas rich systems
%in NG groups.

In this work, we propose to investigate the evolution of cluster galaxies 
using the plane defined by the mean stellar age (hereafter, ``Age") and specific star formation rate (hereafter, ``SSFR") of galaxies to define their evolutionary stages. In \S 2 we define the cluster sample and describe their properties. Classifications in the Age-SSFR space and comparison of galaxy properties are performed in \S 3. In \S 4 we revisit the VDP problem, now splitting profiles according to our evolutionary classifications. Also in \S 4 we study the asymmetries of infalling galaxies onto
G and NG systems. Finally, in \S 5 we summarize our results. Throughout the paper, were assumed the $\Lambda$CDM cosmology: H$_0$=72 km s$^{-1}$Mpc$^{-1}$,  $\Omega_{\textrm{m}}$=0.27 and $\Omega_{\Lambda}$=0.73. 

\section{Data}

Our sample of galaxy clusters is defined from the \cite{yang07} (hereafter Y07) updated galaxy group catalog, constructed using a halo-based group finder optimized for grouping galaxies residing in the same dark matter halo \citep{2005MNRAS.356.1293Y}. 

\subsection{General description}

We briefly describe the data selection, which is the same as described by \citetalias{decarvalho} and \citet{costa17}. Galaxies were retrieved galaxies from the Sloan Digital Sky Survey \citep[SDSS;][]{york2000} Data Release 7 within $\pm4000$ km/s and up to 3 Mpc in projected distance around 344 Y07 group center coordinates, limited to $0.03<z<0.1$ and Petrosian magnitude $r<17.77$. Galaxy membership was determined by the \textit{shiftgapper} iterative technique \citep{1996ApJ...473..670F, 2009MNRAS.399.2201L}, and relevant dynamical properties such as velocity dispersion, ${\rm R_{200}}$ and ${\rm M_{200}}$ were determined from the virial analysis \citep[see details in][]{2009MNRAS.399.2201L}.
After the interloper elimination process, we consider two subsamples pertaining to different luminosity regimes. First we selected galaxies within ${\rm 0.03\leq z\leq 0.04}$ and ${\rm -20.5 < M_r\leq -18.4}$, which will be denoted as \textit{faint} sample (F). The second subsample comprises galaxies with ${\rm 0.04<z\leq0.1}$ and ${\rm M_r\leq-20.5}$, thus defining the bright sample (B). Some additional information were added from other available catalogs, which we briefly describe in the next paragraphs.

\subsection{Additional information}

We add some spectroscopic properties from SDSS tables performing a cross match by coordinates and inside a $0.5^{\prime\prime}$ radius, using the CasJobs\footnote{\url{http://skyserver.sdss.org/CasJobs/}} platform. Thus, we retrieve the stellar masses and SSFRs for our sample from the \textit{galSpecExtra} table \citep{2004MNRAS.353..713K} corresponding to the \textit{lgm\_tot\_p50} and \textit{specsfr\_tot\_p50} parameters, respectively.
A total of 84 objects without SSFR and/or age reliable
measurements were discarded, and they represent only a small
percentage (1.2\%) of the whole sample.
We also include the eclass spectral classification from the \textit{SpecObj} table in the CAS, which is based on the first two expansion coefficients of the PCA decomposition for 170,000 SDSS spectra \citep{Yip04}. The eclass parameter ranges from -0.35 to 0.5 for early- to late-type galaxies or AGN. Luminosity-weighted mean stellar ages and metallicities used in this work come from \citetalias{decarvalho}, derived using the STARLIGHT code \citep{cid05}. The errors of both stellar population parameters is assessed by \citetalias{decarvalho} in two different ways. First, the estimates are compared to those of \citet{chen12}, based on  Principal Component Analysis (PCA), and giving the following residuals relative to SDSS-DR12 values: $\Delta$Age$=3.5\pm 2.7$ Gyr; $\Delta$[Z/H]$=-0.01\pm 0.08$. The second way is via repeated observations of the same galaxy from SDSS-DR7 (6148 repeated observations of 2543 galaxies), yielding a residual distribution with the following uncertainties: $\Delta$Age$=0.0\pm 1.2$ Gyr; $\Delta$[Z/H]$=0.00\pm 0.04$ (see more details in \citetalias{decarvalho}). 

We include morphological information for galaxies in our sample in two ways. First, we use the \cite{dominguez2018improving} catalog, which provides T morphological types for $\sim 670,000$ galaxies from SDSS, by training Convolutional Neural Networks (CNNs) with information from available sources such as Galaxy Zoo 2 \citep[GZ2,][]{2008MNRAS.389.1179L,2013MNRAS.435.2835W} and the catalog of visual classifications provided by \cite{nair2010catalog}. Secondly, we include the morphometric parameter G2, from the catalog of \cite{2020A&C....3000334B}. This parameter  is obtained applying the \textit{CyMorph} algorithm on GZ2 galaxies, that uses the second gradient moments
of the galaxy image to compute G2 \citep{rosa18}.
We also add bulge, disk, and total stellar mass from the catalog of \citet{mendel2013catalog} for 
$\sim 660,000$ galaxies from the Legacy area of the SDSS DR7. These masses are based on a homogeneous catalog of 
g and r band photometry described by \citet{simard2011catalog} and extended with bulge$+$disk and S\'ersic profile photometric decompositions in the SDSS u, i, and z bands.
We also define a field sample, comprising 5997 galaxies, which is a result of stacking three catalogs available in the literature \citep{khim2015demographics, bradford2015study, fernandez2015isolated} in the SDSS DR7, DR8 \& DR10 fields, respectively, and removing repeated objects and those belonging to galaxy clusters catalogs of \cite{yang07} and \cite{tempel2017merging}.

With all added information, we have a quite general characterization of the galaxies in our sample, in both photometry and spectroscopy. 
%we used the \cite{dominguez2018improving} catalog, which provides T morphological types for $\sim 670,000$ galaxies from SDSS, by training Convolutional Neural Networks (CNNs) with information from available sources such as Galaxy Zoo 2 \citep[GZ2,][]{2008MNRAS.389.1179L,2013MNRAS.435.2835W} and the catalog of visual classifications provided by \cite{nair2010catalog}.
%\renewcommand{\arraystretch}{1.2}
%\setlength{\tabcolsep}{5pt}

\begin{table}
	\centering
\caption{Properties of the four subsamples to be studied. Median errors were determined as ${\rm 0.7415(Q_{0.75}-Q_{0.25})}$, where ${\rm Q_{0.25}}$ and ${\rm Q_{0.75}}$ are 25\% and 75\% probability quantiles, respectively.}
	\begin{tabular}{ccccc}
		\hline
		Sample & N$_\textrm{cl}$ & N$_\textrm{gal}$ & ${\rm \log(M_{200}/M_\odot)}$ & ${\rm R_{200}(\textrm{Mpc})}$ \\ 
		\hline
		 GB & 98 & 3662 & 14.45 $\pm$ 0.30 & 1.10 $\pm$ 0.25 \\
		 NGB & 26 & 1239 & 14.67 $\pm$ 0.53 & 1.23 $\pm$ 0.56 \\
		 GF & 12 & 699 & 14.38 $\pm$ 0.24 & 0.95 $\pm$ 0.19 \\
		 NGF & 10 & 1042 & 14.74 $\pm$ 0.31 & 1.32 $\pm$ 0.32 \\ 
		  \hline
	\end{tabular}
\label{tab1}
\end{table}	

%\begin{table}
%	\centering
%\caption{Properties of the four subsamples to be studied. Median errors were determined as $0.7415(Q_{0.75}-Q_{0.25})$, where $Q_{0.25}$ and $Q_{0.75}$ are 25\% and 75\% probability quantiles, respectively.}
%	\begin{tabular}{ccccc}
%		\hline
%		Sample & N$_\textrm{{cl}}$ & N$_\textrm{{gal}}$ & $\log(M_{200}/M_\odot)$ & $R_{200}(\textrm{Mpc}\;{\rm h^{-1}})$ \\ 
%		\hline
%		 GB & 98 & 3662 & 14.45 $\pm$ 0.30 & 1.10 $\pm$ 0.25 \\
%		 NGB & 26 & 1239 & 14.67 $\pm$ 0.53 & 1.23 $\pm$ 0.56 \\
%		 GF & 12 & 699 & 14.38 $\pm$ 0.24 & 0.95 $\pm$ 0.19 \\
%		 NGF & 10 & 1042 & 14.74 $\pm$ 0.31 & 1.32 $\pm$ 0.32 \\ 
%		  \hline
%	\end{tabular}
%\label{tab1}
%\end{table}	

\subsection{Dynamical state of clusters}

Since we want to relate the galaxy properties with the dynamical state of the clusters, we add the classifications find by \citetalias{decarvalho}, into Gaussian or non-Gaussian. \citetalias{decarvalho} assess the Gaussianity based on the Hellinger Distance (HD) \citep[e.g.][]{huber1981robust,amari1985differential} estimator over the cluster velocity distribution, establishing a reliability threshold depending on the cluster richness (see details in \citetalias{decarvalho}). HD is a measure of the distance between two distributions, first introduced in astronomy by \cite{2013MNRAS.434..784R}, who demonstrate that HD overcomes other normality tests when comparing the amount of type I and II statistical error rates. Furthermore, \citetalias{decarvalho} compare HD with \textit{mclust}, an independent method (see description in Section 3.1), obtaining similar results, although HD was find to perform slightly better. HD reliability is achieved for a richness of, at least, 20 galaxies, which translates in an additional total mass selection for our sample of ${\rm M>10^{14}M_{\sun}}$ (see \citetalias{decarvalho}). This richness cut reduces the initial sample from 344 to 146 clusters, containing 6642 galaxies.
In this work, we compare the properties of galaxies between Gaussian (G) and non-Gaussian (NG) systems, separating by luminosity regime Bright (B) or Faint (F), and thus defining the four subsamples to be analyzed subsequently: GB, NGB, GF and NGF. Table \ref{tab1} shows the main properties of the defined subsamples.

\section{Classification and evolution of cluster galaxies}

Although colors is widely used to identify populations of different degrees of star formation, it is well known that they suffer from degeneracy in age, metallicity and dust extinction \citep[e.g.][]{gallazzi05}, and cannot represent low values of star formation rate \citep[see e.g.][]{dressler15}. \citet{cortese2012passive}, using UV and mid-IR star formation rates, conclude that optical colors alone are not able
to distinguish between actively star-forming and passive galaxies.
The use of ${\rm NUV - r}$ selection \citep[e.g.][]{wyder07} has the advantage of being model-independent
at $z \approx 0$, but even at very low redshifts it remains
difficult to correct the NUV flux for the effects of dust extinction in
absence of detailed multi-band photometry \citep{belfiore2018sdss}. Besides, using UV $-$ optical colors always reduces the sample size.
In the case of our sample this reduction reaches $\sim22$\%, since not all sources of SDSS have been observed by the \textit{Galaxy Evolution Explorer} (GALEX). 

Some authors overcome the color issue using the SSFR instead \citep[e.g.][]{2012MNRAS.424..232W}, which can show more sharply the degree of quiescence and galaxy bimodality. Furthermore, SSFR is a good proxy for the star formation history and the evolutionary stage of galaxies \citep[e.g.][]{salim14,brennan15,eales17}. 
In this work, in addition to SSFR, we also use the light-weighted age to characterize galaxy evolution.

\subsection{The Age-SSFR plane}

\begin{figure}
	\centering
	\includegraphics[width=1\linewidth]{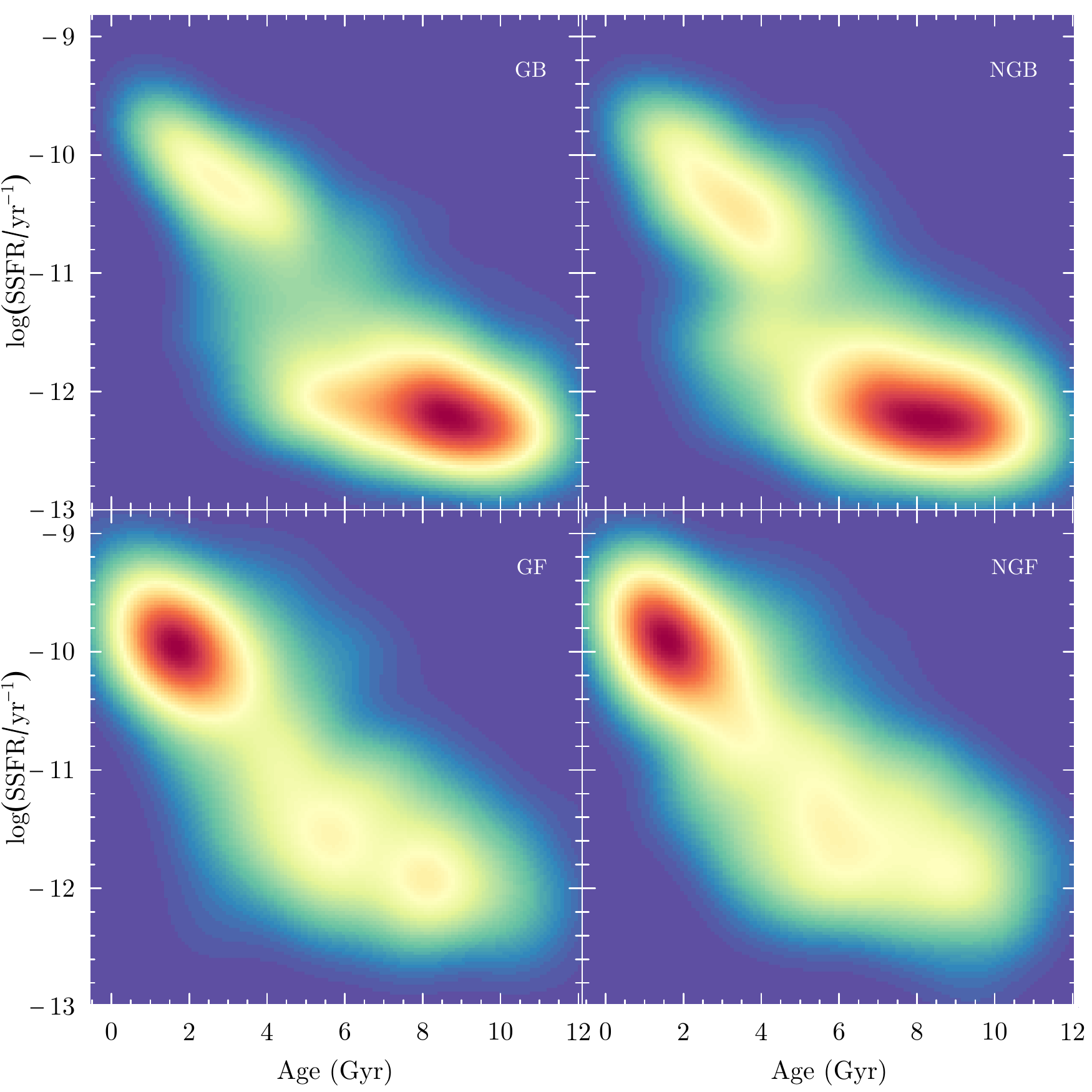}
	\caption{Density maps for GB, NGB, GF and NGF samples on the Age-SSFR plane. The density increases with color, from blue to red. }
%The horizontal dashed lines divide the active/passive subsamples according to $\log(\textrm{SSFR}/\textrm{yr}^{-1})=-11$ \citep{2012MNRAS.424..232W}. The vertical dashed lines indicate the blue cloud as objects with ${\rm Age}<2.5$ Gyr \citet{thomas2010environment}.}
\label{fig1}
\end{figure}

\begin{figure}
	\centering
	\includegraphics[width=1\linewidth]{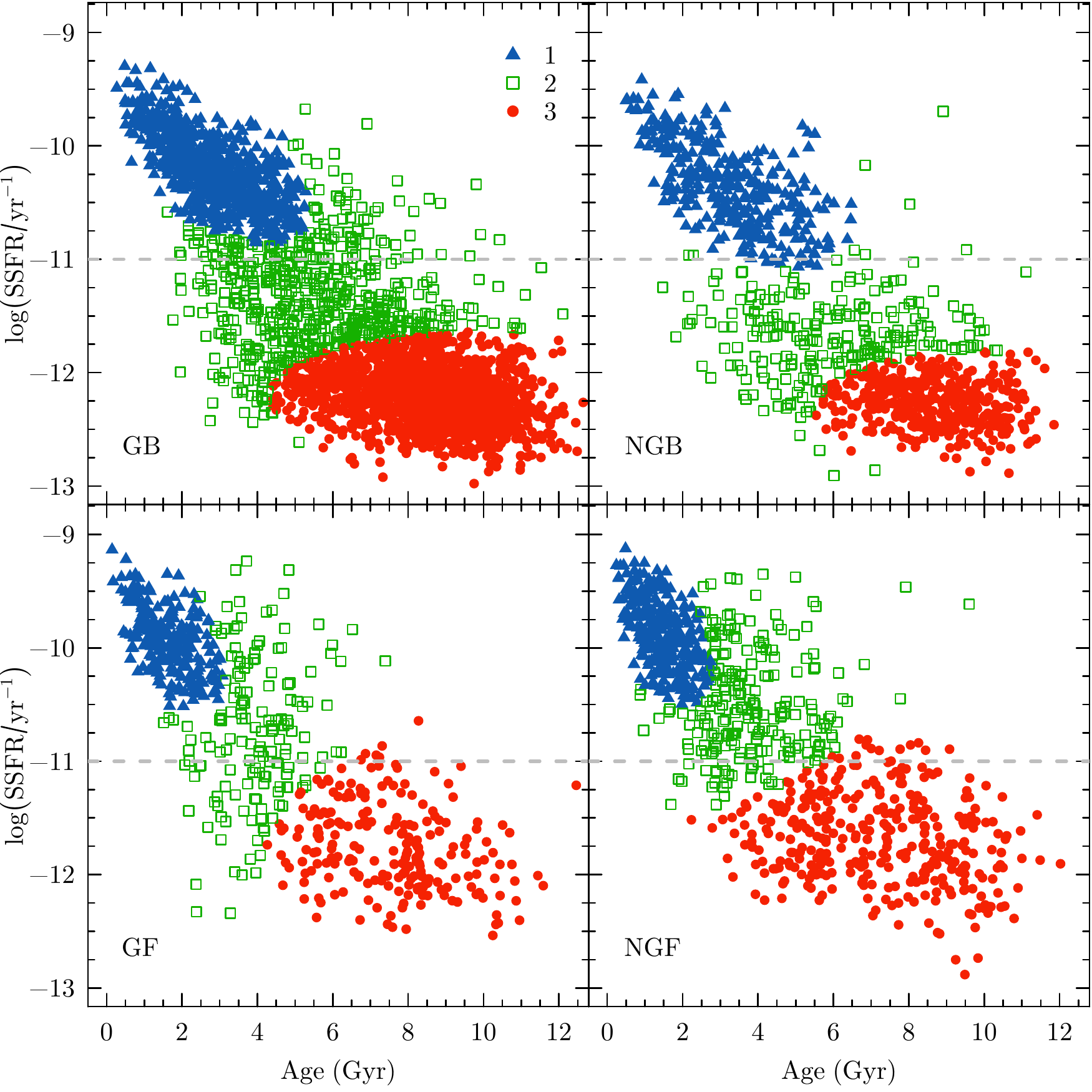}
	\caption{Classification given by \textit{mclust} for 3 components in the Age-SSFR space, for each subsample. The horizontal dashed lines indicate the usual separator
of active and passive galaxies given by \citet{2012MNRAS.424..232W}.}
\label{fig2}
\end{figure}

SSFR and mean stellar age are both bimodal properties of galaxies at low and medium redshifts \citep[e.g.][]{2003MNRAS.341...33K,gallazzi05,2012MNRAS.424..232W}. The former indicates the current star formation normalized by galaxy mass, and the latter is a tracer of the time elapsed since the last burst of star formation. 
Galaxies following the star forming main sequence (SFMS) are typically found at lower masses, while very low SSFR dominate the high-mass end. This behavior is often interpreted as galaxies growing along the SFMS at a rather constant SFR until they reach a critical stellar mass (or halo mass) threshold, above which they likely quench \citep[e.g.][]{peng10,abramson14}. Considering that 
stellar mass represents the past star formation, then
SSFR compares the current star formation to the averaged past one. 
If one assumes a constant SFR with time at low redshifts \citep[e.g.][]{kennicutt1998star} then SSFR scales as the inverse of the age of the galaxies. In the nearby universe the SSFR decreases when stellar mass increases, i.e., massive galaxies formed their stars in the past whereas less massive ones are more active in star formation, or equivalently, the mean age of the stellar populations of massive galaxies is higher than that of low mass galaxies \citep[e.g.][]{li2011dependence,belfiore2018sdss}.
This suggests a bimodal distribution with well separated peaks 
on the Age-SSFR plane, providing us a useful tool for classifying galaxies into evolutionary groups.

%\begin{figure}
%	\centering
%	\includegraphics[width=1\linewidth]{nsfms.pdf}
%	\caption{Star forming main sequence for GB+NGB+GF+NGF samples in the SFR versus
%$M_\ast$ diagram. colors indicate groups
%from {\textit mclust} classification. The SFMS plus 1$\sigma$ around its linear fit is shown.}
%\label{fig2} 
%\end{figure}

In Figure \ref{fig1} we show the distributions of the four subsamples defined in the previous section (GB, NGB, GF, NGF) on this plane.
In all panels  we verify the expected bimodality, with well separated peaks with higher concentrations of old and low-SSFR objects in samples GB and NGB, with an inversion of this behavior in samples GF and NGF, as expected in the downsizing scenario \citep[e.g.][]{neistein2006natural,fontanot2009many,gu2018morphological}.
Our first aim is to classify objects in this space. This is done using the \textit{mclust} R code \citep{mclust}, that has been used successfully in astronomy  \citep[e.g.][dC17]{2012arXiv1210.1751E,2013MNRAS.434..784R}. 

In model-based clustering, data is considered a mixture of densities. 
Each component in the mixture is modeled by the normal distribution which is characterized by its mean and covariance matrix.
Briefly, \textit{mclust} performs clustering by fitting Gaussian finite mixture models to the data, using an expectation-maximization algorithm, and further giving the number of optimal components via the lower Bayesian information criterion (BIC). 
We apply \textit{mclust} to the whole sample choosing the more general covariance model ``VVV" (varying volume, shape, and orientation), which returns the best gaussian mix being the one with three components. Considering that this is exactly the number of components we would intuitively expects (``blue", ``red" and ``green" family of galaxies) we test the robustness of such a finding. For each subsample, we run bootstrap realizations of the plane $\rm SSFR \times \rm Age$ one-hundred times fixing one quantity at a time. Fixing Age we find an outcome of three components 83\% of the time and fixing SSFR, 91\%. The subject of ``clustering" in 2D
is certainly a critical and important one and goes beyond the scope of the present investigation. However, the test we apply shows a consistency that matches
the current vision of how galaxies evolve in the  ``$\rm SSFR \times \rm Age$" plane.

%We applied \textit{mclust} to the whole sample choosing the more general covariance model ``VVV" (varying volume, shape, and orientation), which returned the best gaussian mix being the one with three components. The code was also run for each subsample and the result is the same in all cases. The result proved to be robust after 1000 resamples in age and SSFR, keeping the diagnostic 83\% and 91\% of the times, respectively.

%Only normal galaxies
%were considered. A total of 84 objects without SSFR and/or Age measurements were removed. We also removed 433
%Seyfert galaxies and 865 LINERS (low ionization nuclear emission-line regions) for further analysis.
%We refine this result by  redefining young objects as those in the star formation region of the BPT diagram \citep{baldwin1981classification,kewley2006host}, and within 1$\sigma$ around the
%SFMS linear fit,  following the work of \citet{belfiore2018sdss}.
%Galaxies lying more than 1$\sigma$ below the SFMS are reclassified as intermediate objects.
%Likewise, intermediate objects within 1$\sigma$ of the linear fit are reclassified 
%as star forming galaxies.

\setlength{\tabcolsep}{1.5pt}
\begin{table}
	\centering
\caption{Counts and median values of SSFR and Age for populations 1, 2, and 3 in the four stacked samples.}
	\begin{tabular}{ccccc}
                \hline
                  ~~  & ~~ & Counts & ~~ & ~~ \\
		\hline
		Population & GB & NGB & GF & NGF \\ 
                  1 & 707 & 330 & 247 & 301 \\
                  2 & 819 & 347 & 179 & 259 \\
                  3 & 2071 & 544 & 265 & 437 \\
               \hline
                ~~ & ~  & $\log(\textrm{SSFR}/\textrm{yr}^{-1})$ & ~~ & ~~ \\
		\hline
                Population & GB & NGB & GF & NGF \\ 
		 1 & -10.2$\pm$0.4 & -10.4$\pm$0.4 & ~-9.9$\pm$0.3 & ~-9.9$\pm$0.3 \\
		 2 & -11.4$\pm$0.4 & -11.7$\pm$0.3 & -10.8$\pm$0.6 & -10.6$\pm$0.5 \\
		 3 & -12.2$\pm$0.3 & -12.3$\pm$0.2 & -11.8$\pm$0.4 & -11.7$\pm$0.4 \\
		  \hline
               ~~  & ~~ & Age (Gyr) & ~~ & ~~ \\
\hline
                 Population & GB & NGB & GF & NGF \\ 
		 1 & 2.9$\pm$1.4 & 3.2$\pm$1.5 & 1.7$\pm$0.8 & 1.5$\pm$0.7 \\
		 2 & 5.4$\pm$1.9 & 5.6$\pm$1.9 & 4.0$\pm$1.0 & 3.7$\pm$1.3 \\
		 3 & 8.5$\pm$1.8 & 8.5$\pm$1.6 & 7.5$\pm$1.8 & 7.1$\pm$2.3 \\
		  \hline

	\end{tabular}
\label{tab2}
\end{table}

%Population & GB & NGB & GF & NGF \\ 
%                  1 & 643 & 238 & 251 & 345 \\
%                  2 & 653 & 321 & 153 & 195 \\
%                  3 & 1495 & 386 & 219 & 344 \\
%		GB & (2.85$\pm$1.36, -10.23$\pm$0.36) & (5.36$\pm$1.91, -11.44$\pm$0.41) & (8.46$\pm$1.78,-12.20$\pm$0.26) \\ 
 %       NGB & (3.23$\pm$1.51, -10.39$\pm$0.44) & (5.64$\pm$1.89, -11.68$\pm$0.28) & (8.50$\pm$1.62, -12.27$\pm$0.22) \\
 %       GF & (1.67$\pm$0.76, -9.94$\pm$0.30) & (3.95$\pm$0.96, -10.81$\pm$0.63) & (7.48$\pm$1.83,  ) \\ 
%        NGF & (1.46$\pm$0.66, -9.89$\pm$0.32) & (3.71$\pm$1.27, -10.60$\pm$0.49) & (7.12$\pm$2.25, -11.68$\pm$0.40) \\ 	

The final
classification groups are numbered as 1, 2 and 3 for blue, green and red points of Figure \ref{fig2}, respectively, where
we also present lines separating star-forming and passive subsamples according to $\log(\textrm{SSFR}/\textrm{yr}^{-1})=-11$ \citep{2012MNRAS.424..232W}.  
Looking at population 3 (red points) we see that the Age range for both bright and faint galaxies in G systems is
the same, extending to $\sim$4 Gyr, but for NG systems this does not happen. While the bright sample 
extends to $\sim$6 Gyr, the faint sample extends to younger ages, $\sim$2 Gyr,
indicating recent episodes of star formation in this type of objects. Still in this figure, note that  galaxies in population 1 (blue points) are distributed in the same way \textit{wrt} luminosity, irrespective of the dynamic state of the host clusters.
The bright samples (G and NG) have population 1 with Age $\lesssim$ 6 Gyr, while the faint samples (G and NG) have younger population 1, with Age $\lesssim$ 3 Gyr.

Counts and median values of SSFR and Age for populations 1, 2, and 3 in the four stacked samples are presented in Table \ref{tab2},
where we see a clear progression of young and high-SSFR objects to older and low-SSFR objects, passing through an intermediate population
of galaxies. This sequence has statistical support from the Kruskal-Wallis (KW) test from the \textit{pgirmess} R package \citep{pgirmess}. The KW test is used to compare two or more populations and tests the null hypothesis that all populations have equal distribution functions against the alternative hypothesis that at least two of the populations have different distribution functions \citep{corder2009comparing}. Importantly, the KW test  corrects non-complementarities, making it very powerful when comparing more than two samples simultaneously.
The test finds significant difference at the 95\% confidence level
between the populations $(1\times 2\times 3)$, irrespective of being G or NG, B or F.
This result indicates that the sequence $1 \rightarrow 2 \rightarrow 3$ can be understood as a physical transition of star forming (SF)
to passive galaxies (PAS), passing through the green valley (GV) \citep{martin2005galaxy}. On the other hand, there seems to be no significant difference between samples G and NG, only between B and F galaxies. Indeed,
from Table \ref{tab2}, we see that the differences in age and SSFR between objects of the same class in G and NG systems are less than the sum of the errors in quadrature around their medians. In the same table, we see  large differences in counts between all samples.

Computing the fractions of galaxies in populations 1, 2 and 3, we find
(20\% $\pm$ 2\%, 23\% $\pm$ 3\%, 57\% $\pm$ 2\%) and (27\% $\pm$ 2\%, 28\% $\pm$ 3\%, 45\% $\pm$ 2\%) for the GB and NGB samples, respectively. Hence, for bright galaxies, the proportion of SF + GV objects is higher in NG systems {\bf (55\% $\pm$ 4\%)} than in G systems {\bf (43\% $\pm$ 3\%)}, and PAS galaxies are dominant in G systems. On the other hand, the proportions find for faint objects are all similar: (36\% $\pm$ 2\%, 26\% $\pm$ 3\%, 38\% $\pm$ 2\%) and (30\% $\pm$ 2\%, 26\% $\pm$ 4\%, 44\% $\pm$ 3\%), for the GF and NGF samples, respectively. This leads us to a first finding: galaxies seem to evolve the same way in all samples, 
following the sequence $1 \rightarrow 2 \rightarrow 3$, but the mixtures of types differ between GB and NGB systems. To better understand this result, 
we need to further investigate the intermediate population, 
and also  use a wider set of properties to evaluate in detail the differences and similarities between the stacked samples.

\subsection{The intermediate population}

\begin{figure}
	\centering
	\includegraphics[width=1\linewidth]{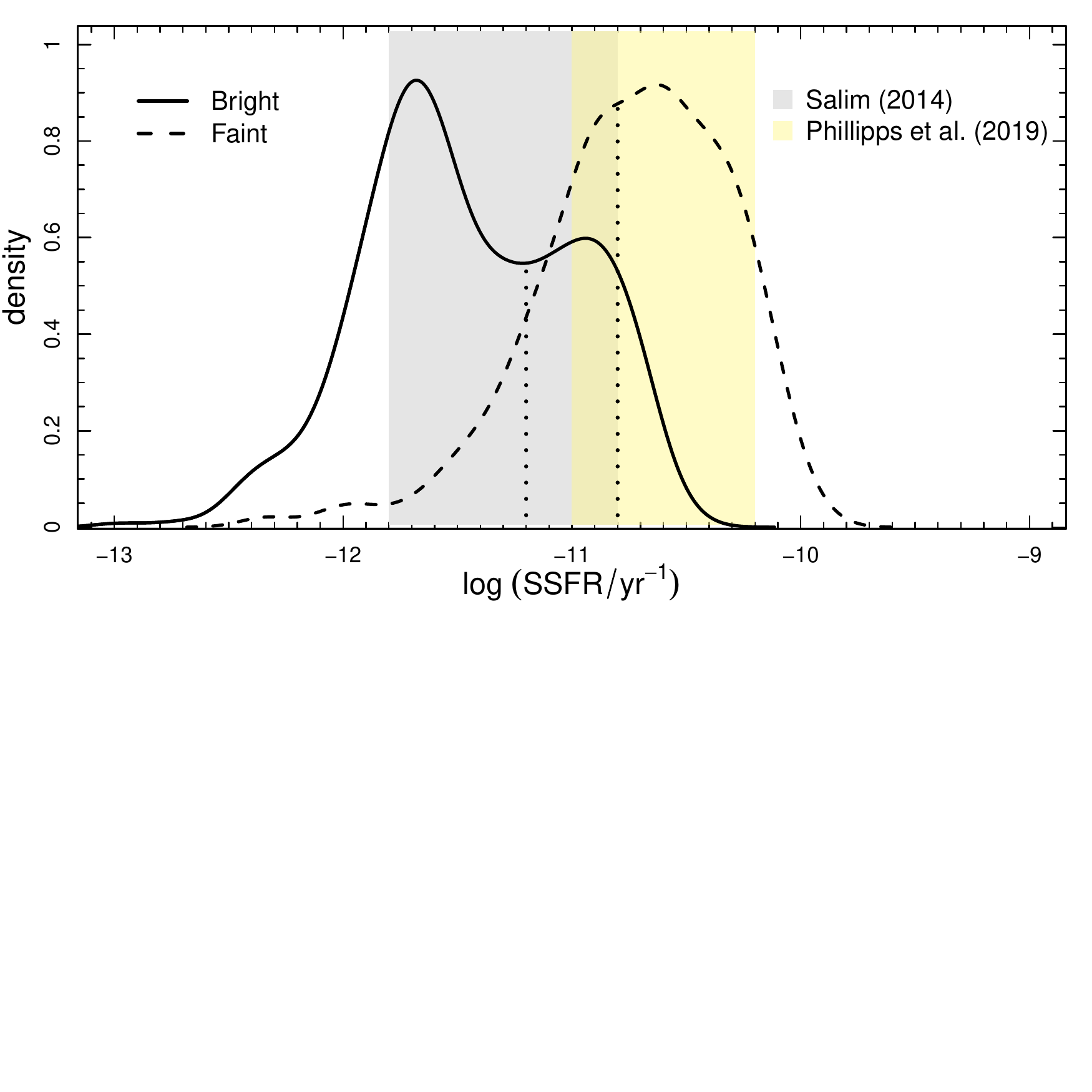}
        \vspace{-4cm}
	\caption{Distributions of SSFR for group 2 objects. Delimiters of GV galaxies from \citet{salim14} and \citet{phillipps2019galaxy} are indicated in gray and yellow, respectively. Vertical dotted lines indicate the separation between passive and star-forming galaxies in each sample.}
\label{fig3}
\end{figure}
\vspace{-0.2cm}

In Figure \ref{fig2}, we note that population 2 consists of a mixture of star-forming and passive objects,
allowing a further characterization of the intermediate galaxies. We also notice that the 
line $\log(\textrm{SSFR}/\textrm{yr}^{-1})=-11$ divides galaxies differently, especially in samples B or F.
In Figure \ref{fig3}, we present the SSFR distribution of bright and faint galaxies only in population 2. 
The B sample presents a clear bimodality while the F sample is a single broad mode.
Typical GV  delimiters
of \citet{salim14} and \citet{phillipps2019galaxy} are shaded in gray and yellow, respectively.
 Note that the B sample is
better delimited by the boundaries of \citet{salim14}, while the F sample is better framed by the delimiters of \citet{phillipps2019galaxy}.
These features indicate
that transitional objects are not homogeneous in luminosity.
Using \textit{mclust}, we
separate star-forming and passive modes of B galaxies at $\log(\textrm{SSFR}/\textrm{yr}^{-1}) \approx -11.2$,
see Figure \ref{fig3}. As a consequence, population 2 is separated 
differently between the samples: 61\% $\pm$ 4\% (39\% $\pm$ 5\%) of passive (SF) for GB; 
and 82\% $\pm$ 4\% (18\% $\pm$ 6\%) for NGB. 
Recall, from Section 3.1, the higher fraction of population 2 galaxies in the NGB sample (33\%) 
than in the GB sample (23\%).
These differences are significant at the 95\% confidence level, according to the \textit{twoSamplePermutationTestProportion} function implemented in the EnvStats R package \citep{EnvStats}, which performs a two-sample permutation test to compare two proportions. 

Regarding F objects, we
split them  at the median, $\log(\textrm{SSFR}/\textrm{yr}^{-1})\approx -10.8$, see  Figure \ref{fig3}. The resultant mixtures have the following proportions:
40\% $\pm$ 5\% of passive and 60\% $\pm$ 4\% of star-forming galaxies for GF; and 22\% $\pm$ 7\% of passive and 78\% $\pm$ 4\% star-forming galaxies for NGF. Hence, faint  galaxies in GV are dominated by star-forming galaxies, but there are more passive objects in the GF sample, and more star-forming objects in the NGF sample. This is the opposite of what happens with bright galaxies.
This result may  be related to the work of
\citet{de2019mass} who find that faint spirals of
NG groups are significantly different from their counterpart
in the G groups, regarding their star formation history.

\subsection{A physically intermediate population?}

\begin{figure}
	\centering
	\includegraphics[width=1\linewidth]{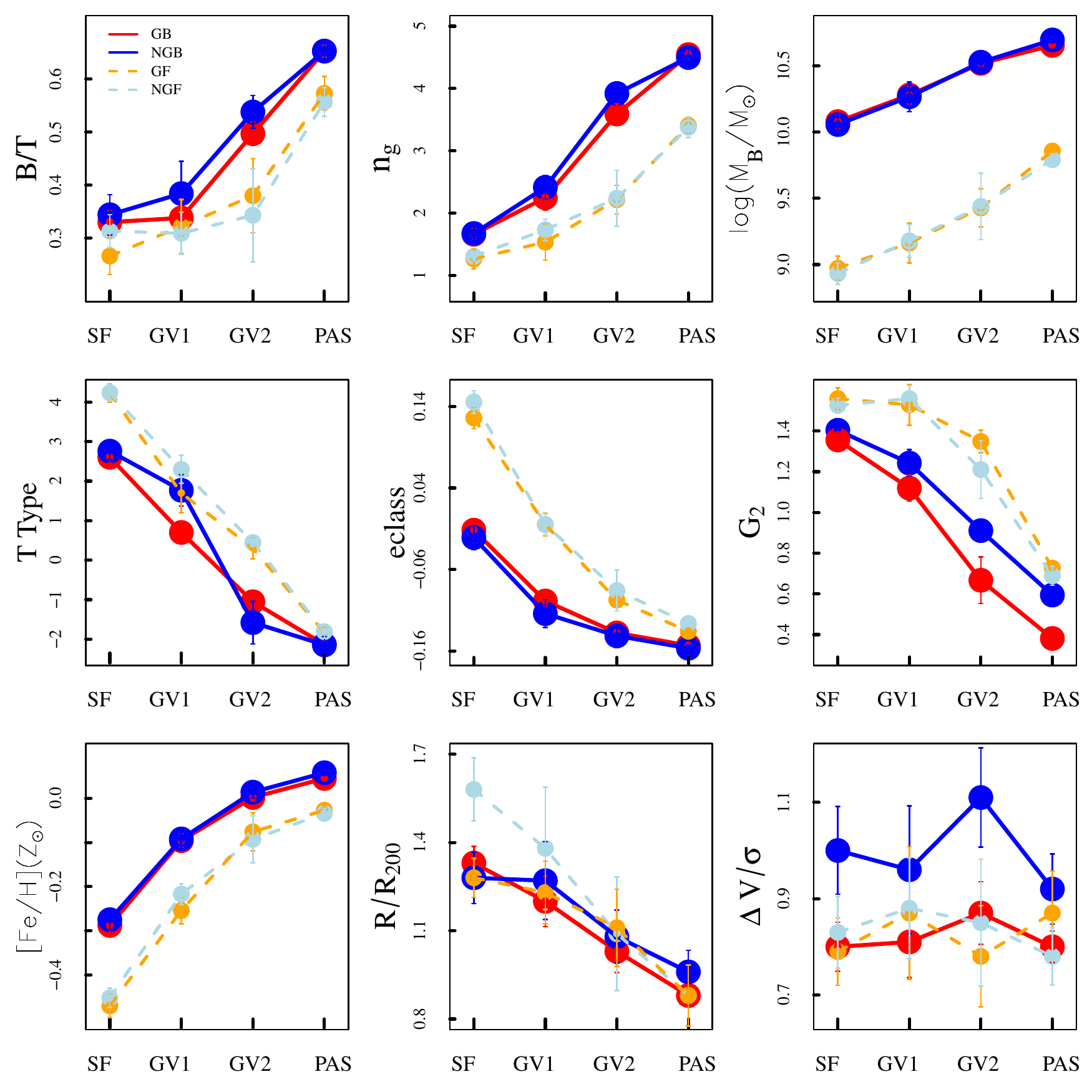}
	\caption{Evolutionary sequence for
properties B/T, ${\rm n_g}$, ${\rm M_B}$, T, eclass, ${\rm G_2}$, [Fe/H], ${\rm R/R_{200}}$, and
${\rm \Delta V/\sigma}$. Samples and tracks are indicated by colors red (GB), blue (NGB), orange (GF),
and light blue (NGF). Solid and dashed lines depict bright and  faint samples, respectively. SF indicates star-forming galaxies, GV1 and GV2 denote the star-forming and passive components of the green valley, and PAS indicates passive objects. Error-bars  are  obtained  from  a  bootstrap technique with 1000 resamplings.}
\label{fig4}
\end{figure}

The results presented in the previous section show that intermediate galaxies correspond to different 
mixtures of passive and star-forming type objects depending on the sample.
It is important to assess whether population 2 is in fact physically intermediate or corresponds to a mixture of types indistinguishable from those of populations 1 and 3, composing a kind of ``purple" population \citep[see][]{mendez11}.
To investigate this point we use a selected set of parameters that allow a more general characterization of our samples.
We first choose parameters linked to the structure of the galaxies (the B/T ratio, the S\'ersic index -- {\bf ${\rm n_g}$}, and the bulge mass, from  \cite{mendel2013catalog}. 
The behaviour of these parameters with respect the sequence ${\rm SF ~\rightarrow GV_{1} \rightarrow GV_{2} \rightarrow PAS}$ is shown in the upper row of Figure \ref{fig4}. GV1 and GV2 denote the star-forming and passive components of the green valley, defined in Section 3.2.
As expected, we see increasing values of all parameters along the sequence. But
note that galaxies in the GV present intermediate values, agreeing with results reported by other authors \citep[e.g.][]{schiminovich2007uv,mendez11,coenda18}.
We also see a significant separation between the trends for objects in samples B and F. Mass and size of objects are likely behind this result, since they increase along the sequence, and thus affect the structural parameters. 

In the middle row, we present the behaviour of three morphological parameters that are defined and computed differently. The well known
T-Type and eclass\footnote{Actually, eclass is the monoparametric spectral classification based on the eigentemplates expansion of galaxy's spectrum, whose values are correlated with galaxy morphology, from about ~ $-0.35$ for early-type galaxies to 0.5 for late-type galaxies \citep{Yip04}.}, plus ${\rm G_2}$. The parameter ${\rm G_2}$ is the second gradient moment
within the GPA (Gradient Pattern Analysis) formalism \citep{rosa1999characterization,andrade2006gradient,rosa18,2020A&C....3000334B}.
After a comparative analysis considering other morphometric parameters, \citet{rosa18} find that 
${\rm G_2}$  is the one with best performance (90\% success) to separate elliptical and spiral galaxies, 
motivating our choice to use it.

Looking at Figure \ref{fig4},
we see that T-Type presents a similar behavior in all samples, except for a few fluctuations.
At the same time, eclass separates samples B and F, with the paths well apart at the beginning of the sequence (SF) and identical at the end (PAS). Finally,
the parameter G2 also shows a separation between samples B and F, but with some differences. First, the paths are less separated at the SF stage. Second, samples GF, NGF and NGB are similar at the end of the sequence (PAS), but note that the behavior is distinct at the GV2 and PAS stages  for galaxies in the GB sample, compared to the NGB sample. This difference means that G systems contain a higher fraction of bright galaxies among the GV2 and PAS populations, than NG systems.

In the lower row of Figure \ref{fig4} we first see metallicity presenting separated  paths 
\textit{wrt} the luminosity of the objects (B $\times$ F), following  the mass-metallicity relation. 
Still in the lower row, we see that the projected clustocentric distance, ${\rm R/R_{200}}$, behaves
the same way for all but one sample, NGF, which presents a more peripheral distribution of SF and GV1 objects.
Finally, we present the behavior of the normalized velocity offsets, ${\rm \Delta V/\sigma}$, which refer to the cluster median velocities (${\rm \Delta V = V_{gal} - V_{cluster}}$), and are scaled by the cluster velocity dispersions (${\rm \sigma}$).  The
velocity offsets have a flat behavior along the evolutionary sequence, although  the NGB sample presents higher velocities, suggesting a distinct kinematic behavior, possibly linked 
to disturbances experienced by NG clusters \citep{2013MNRAS.434..784R,decarvalho}.

\begin{figure}
	\centering
	\includegraphics[width=1\linewidth]{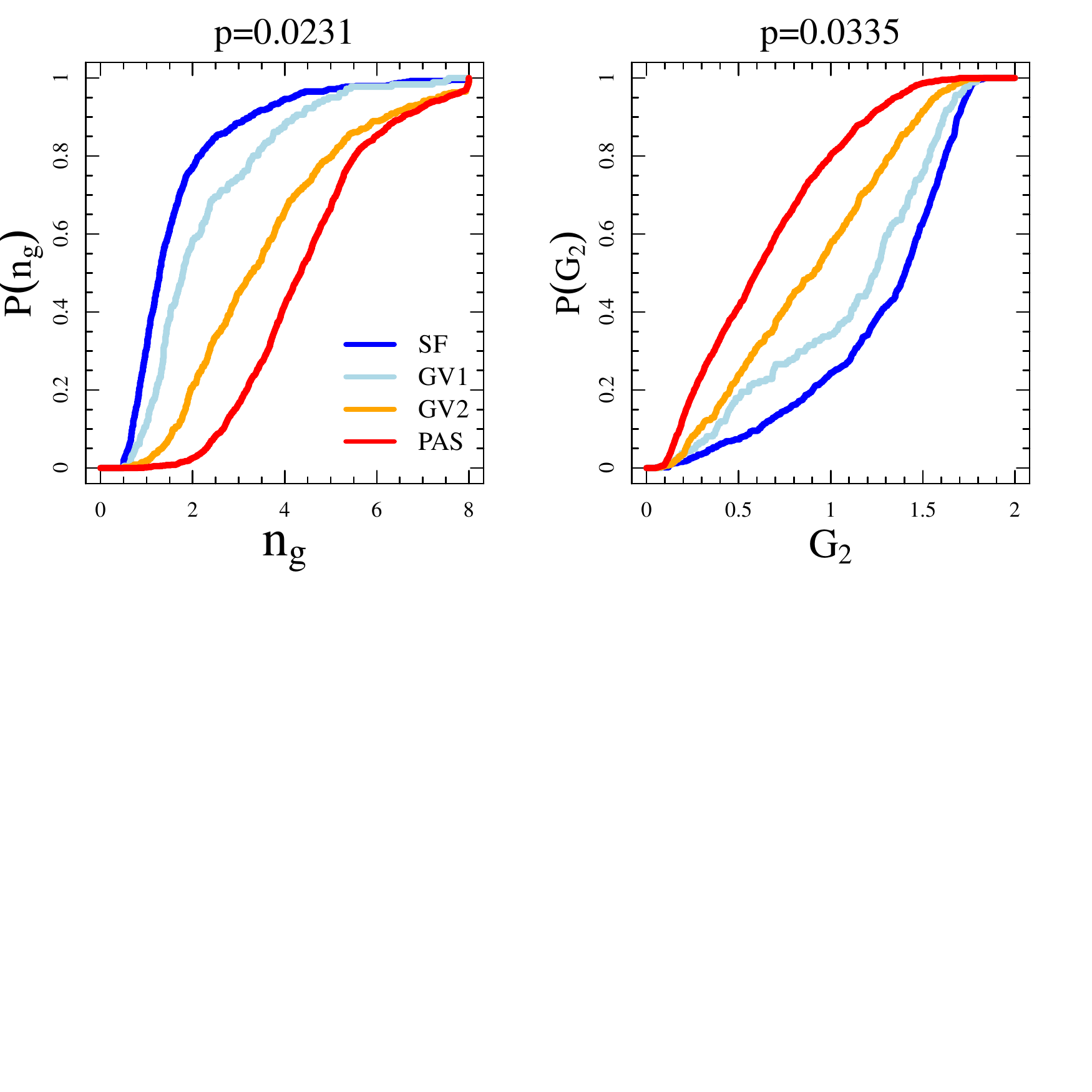}
\vspace{-4.2cm}
	\caption{Example of the Conover test diagnostics for the cumulative distribution function of the
S\'ersic index and the G2 parameter for galaxies in GB systems. Curves are indicated by colors red (GB), blue (NGB), orange (GF),
and light blue (NGF).}
\label{fig5}
\end{figure}

Except for velocities and B/T, all other properties in Figure \ref{fig4} have significant differences between one stage of evolution and another (within each sample).
This has the statistical support of 
the Conover test \citep{conover1999several}, from the \textit{DescTools} R package \citep{signorell2016desctools}.
This test performs a multiple comparison between the datasets and verifies whether the cumulative distribution function (CDF) of one does not cross  the CDF of the other at the 90\% c.l. As an example, we show in
Figure \ref{fig5} the  Conover test diagnostics for the cumulative distribution function of the
S\'ersic index, ${\rm n_g}$, and the G2 parameter for galaxies in GB systems. Note in this figure that
curves for GV1 and GV2 are not coincident with those for SF and PAS, respectively.
The outcome not only reinforces the idea of an evolutionary sequence, but it also
means that SF galaxies are different 
from GV1 and PAS objects are distinct from GV2, for seven (of nine) properties studied in this work. This result may indicate that galaxies in GV are not just a ``purple" mixture of bulges and disks, but actually correspond to objects at an intermediate stage of their evolution.

\subsection{Morphological trends}

\begin{figure}
	\centering
	\includegraphics[width=1\linewidth]{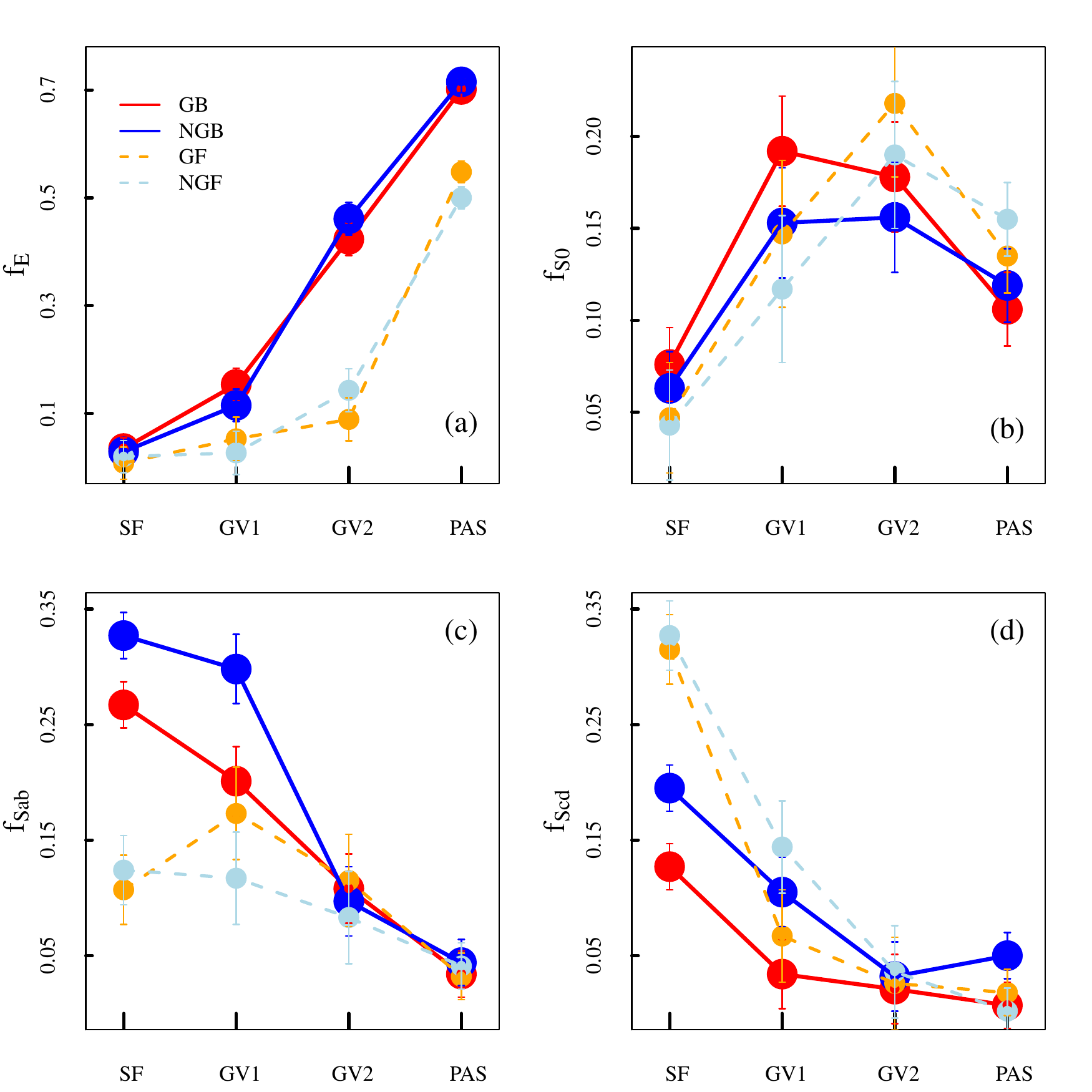}
	\caption{Fractions of galaxy types
E, S0, Sab, and Scd along the evolutionary sequence ${\rm SF ~\rightarrow GV_{1} \rightarrow GV_{2} \rightarrow PAS}$. Samples and tracks are indicated by colors red (GB), blue (NGB), orange (GF),
and light blue (NGF). Solid and dashed lines depict bright and  faint samples, respectively. SF indicates star-forming galaxies, GV1 and GV2 denote the
active and passive components of the green valley, and PAS indicates passive objects.}
\label{fig6}
\end{figure}

In the previous section, we find a small (but significant)  morphological difference between GB and NGB systems at the GV2 and PAS stages, according to parameter G2. 
This indicates that G systems contain a higher fraction of early-type galaxies in the GV2 and PAS populations compared to NG systems, or could mean that morphological transformation times are shorter in G clusters for bright galaxies.
This  can be better understood if we study how morphological types are distributed in these systems.
Using information from \cite{dominguez2018improving} catalog, we present in Figure \ref{fig6} the fractions of galaxy types
E, S0, Sab, and Scd along the evolutionary sequence introduced in the previous sections. In this analysis, we remove all objects
without reliable classification, which represent $\sim$19\% of the whole sample.\footnote{The removal does not significantly modify the distribution of galaxy types defined in Section 3.1 among the four samples under study.}

In panel (a) of Figure \ref{fig6}, we see
the expected increasing of the elliptical fraction in all samples, with  different paths for B and F samples, and a high fraction of ellipticals allocated in GB and NGB systems at GV2 bin, compared to GF and NGF systems.  The result indicates a dichotomy in the formation of ellipticals \textit{wrt} their luminosities.  We draw attention to the transition GV1 $\rightarrow$ GV2, much more pronounced for bright galaxies, revealing the importance of transformations through the GV to separate the paths of bright and faint objects. 

On panel (b),
we  see that galaxies S0 types are more frequent among GV galaxies than among SF and PAS galaxies. Actually, GV1 is dominated by S0 and Sab types, while GV2 is dominated by S0 and E,
considering all the panels in the figure.
This suggests evolution through the GV, and that S0
type is crucial for understanding the transition from late to early type objects, in agreement with recent works \citep[e.g.][]{bait2017interdependence, bremer18}. 
At the same time, the fraction of Sab
decreases from SF to PAS. It should be noted, on panel (c), that the Sab fraction is higher in bright samples at the SF and GV1 bins, and that the NGB sample has the highest fraction of these objects. Indeed,
the fraction of Sab galaxies in NGB systems,
${\rm f_{Sab}^{NGB}}$, has a small drop over the ${\rm SF \rightarrow GV1}$ transition, and a significant drop during ${\rm GV1 \rightarrow GV2}$. This is accompanied by a significant growth in 
the fraction of S0 objects in these systems,
${\rm f_{S0}^{NGB}}$, over ${\rm SF \rightarrow GV1}$, and just a small growth fraction over ${\rm GV1 \rightarrow GV2}$. For galaxies in  GB systems we see a similar behaviour, but note that the decrease in
the fraction of Sab galaxies in GB systems,
${\rm f_{Sab}^{GB}}$, during the same transitions (${\rm SF \rightarrow GV1}$ and ${\rm GV1 \rightarrow GV2}$), is lower, suggesting that environmental effects are somehow regulated by the dynamic state of the clusters. Still in panel (c), we see that there is no significant reduction of faint objects 
over ${\rm SF \rightarrow GV1}$, and only a modest decrease over ${\rm GV1 \rightarrow GV2}$. This suggests that, in the case of low luminosity galaxies, Sab objects do not decrease enough to achieve the S0 growth observed in panel (b).
This brings us to panel (d), where we see the impressive decrease of faint Scd objects during the aforementioned transitions, indicating that the S0 growth in F samples is correlated with the depletion of Scd galaxies both in G and NG systems.
We still highlight in panels (c) and (d) an important difference in the proportion of types Sab and Scd at the SF bin. While the proportion of Sab objects is much higher for bright galaxies, the proportion of Scd objects is much higher for the faint ones, indicating that the bulk of the star forming main sequence corresponds to Sab or Scd types, depending on the luminosity range of the
sample.

As a general result, the fractions of galaxies E increases while those of Sab and Scd decrease along the
sequence ${\rm SF ~\rightarrow GV_{1} \rightarrow GV_{2} \rightarrow PAS}$.
Reducing a population means transformation or destruction of individual objects. Increasing a population means receiving a new contingent or having other populations within the clusters being converted to that one. But new contingents via infall do not seem sufficient to recover the Sab and Scd galaxy counts. Also, recall that GV1 is dominated by a mixture of S0 and Sab, while GV2 is more populated by S0 galaxies (and E galaxies in GB and NGB samples). At the same time, there is a significant increase in S0 counts concomitant with a drop in Scd objects for faint galaxies from SF to GV2. Finally,
ellipticals dominate the PAS subsample with a drop of S0 objects. Despite the details of this morphological  balance,  these facts suggest the GV as the stage of galaxy evolution where the transition from types Sab and Scd to S0 must be taking place, and that some S0 are converted into ellipticals. This view agrees with the scenario proposed by \citet{bait2017interdependence}, where the fraction of early-type spirals decreases towards the green valley from the blue cloud, coinciding with the increase in the fraction of S0 due to environmental effects. Our results add to this scenario that the fraction of faint late-type spirals also decreases accompanied by an increase of the fraction of faint S0 galaxies.

\begin{figure}
	%\centering
%\raggedleft 
%\begin{flushright}
	\includegraphics[scale=0.475]{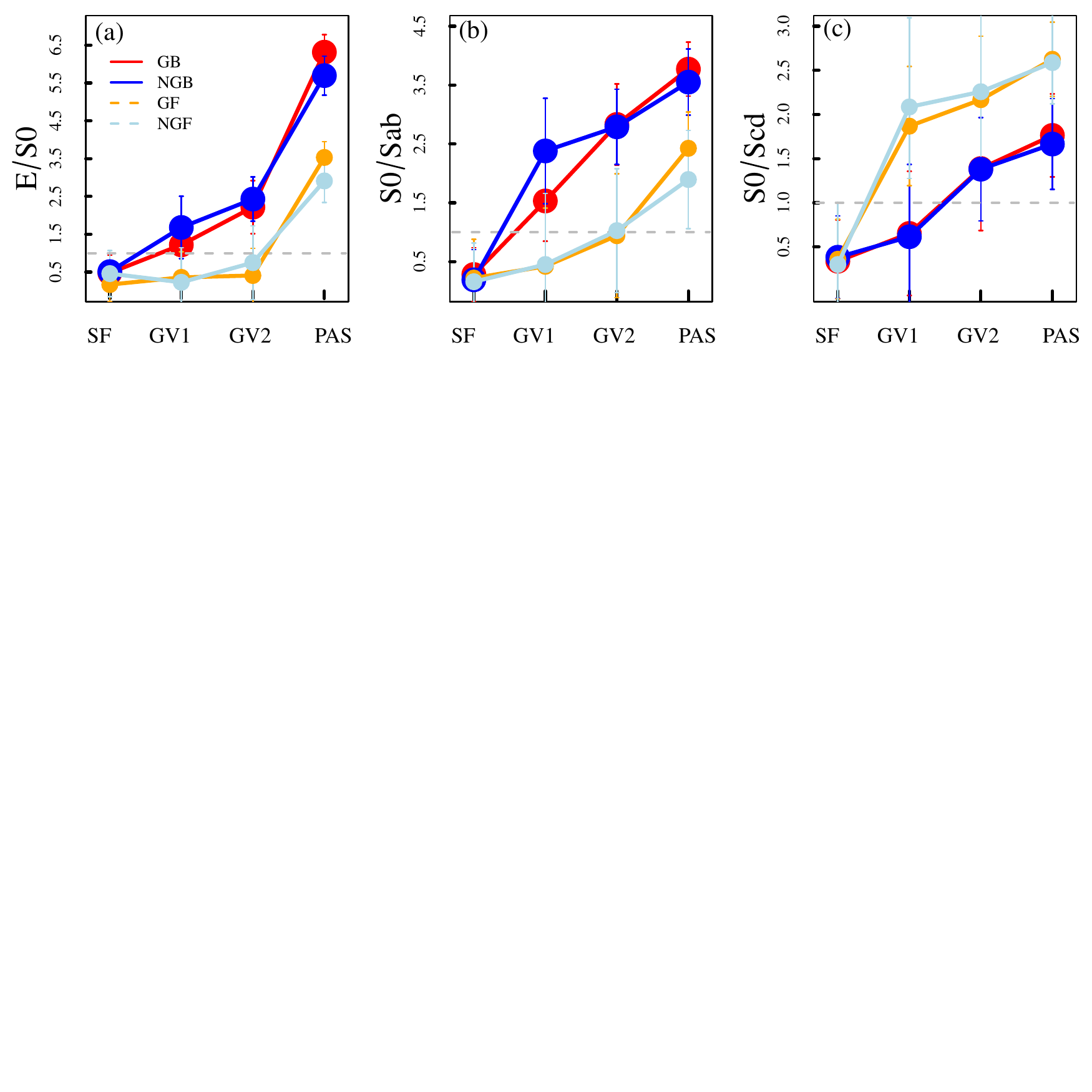}
%\end{flushright}
\vspace{-5.8cm}
	\caption{Number ratios E/S0, S0/Sab, and S0/Scd along the evolutionary sequence ${\rm SF ~\rightarrow GV_{1} \rightarrow GV_{2} \rightarrow PAS}$. Samples and tracks are indicated by colors red (GB), blue (NGB), orange (GF),
and light blue (NGF). Solid and dashed lines depict bright and  faint samples, respectively. SF indicates star-forming galaxies, GV1 and GV2 denote the
active and passive components of the green valley, and PAS indicates passive objects. Dashed lines indicate equal counts of the morphological types.}
\label{fig7}
\end{figure}

In Figure \ref{fig7} we reinforce this view presenting the number ratios of morphological types E/S0, S0/Sab, and S0/Scd. 
On panel (a) we see that the E/S0 ratio is increasing along the sequence, but notice that the growth rate is higher for bright galaxies,
for which the ratio ${\rm E/S0 > 1}$ at the GV1 stage, while for faint galaxies this only happens after GV2.
This suggests that the ${\rm S0~ \rightarrow ~E}$ conversion is regulated by galaxy luminosity.
On panel (b) we see a similar behavior regarding the conversion ${\rm Sab~ \rightarrow ~S0}$. In this case, note that the 
${\rm S0/Sab > 1}$ during the ${\rm SF~ \rightarrow ~GV1}$ transition for bright galaxies, while for faint galaxies this happens at the GV2 stage. Finally, we see in panel (c) that ${\rm S0/Scd > 1}$ over the ${\rm SF~ \rightarrow ~GV1}$ transition, for faint galaxies, while
this happens over ${\rm GV1~ \rightarrow ~GV2}$, for bright galaxies. We highlight the strong dependence on galaxy luminosities regarding the conversions between morphological types.
It is worth mentioning that conversion of one morphological type into another seems to be independent of the dynamical
stage of the clusters, namely, the behaviors for E/S0, S0/Sab, and Scd are similar for G and NG systems.

%\section{Radial trends}

%Weights were assumed for the conversion (Sab + Scd) into S0, and defined as being proportional to the respective counts in each sample.

%\begin{figure}
	%\centering
%\raggedleft 
%\begin{flushright}
%	\includegraphics[scale=0.475]{radprop.pdf}
%\end{flushright}
%\vspace{-0.5cm}
%	\caption{Radial profiles of SSFR, G2 and ${\rm n_g}$. Colors and lines follow the same definitions as the previous plots.}
%\label{fig7}
%\end{figure}

%Therefore, the analysis of velocity dispersion as a function of the clustocentric radius allows us to assess the kinematic youth of galaxies and help us to clarify the picture of evolution in cluster environments.
%We analyze the kinematics by comparing the radial velocity dispersion profiles (VDPs) for the evolutionary classes in each subsample. 

%\begin{figure}
	%\centering
%\raggedleft 
%\begin{flushright}
%	\includegraphics[scale=0.475]{vdisp2.pdf}
%\end{flushright}
%\vspace{-4.5cm}
%	\caption{Radial and stellar mass profiles of the normalized velocity dispersion ${\rm \sigma_u}$. Colors and lines follow the same definitions as the previous plots.}
%\label{fig8}
%\end{figure}

\section{Velocity Dispersion Profiles}

%The cluster dynamical state is related to the orbits of member galaxies. As a galaxy moves through the cluster, its orbit is eventually modified, from highly radial trajectories in the cluster outskirts to circular orbits in central regions inside the virial radius. Along with galaxy dynamical history, physical properties also changes influenced by the cluster environment, and so can be linked to the orbital type. For example, different Hubble types are associated with different orbits, with early-type galaxies having more isotropic orbits than late-type galaxies \citep[e.g.][]{tammann72,sodre89,adami98,biviano2002eso,aguerri07,cava17}. 
%The VDP shape is also related to some properties such as the efficiency of merger activity and/or substructure \citep[e.g.][]{menci96,2012MNRAS.421.3594H,pimbblet14,bilton18}, presence of different spectral classes galaxies \citep{rood72} and cluster dynamical state \citep[e.g.][]{hou2009statistical,costa17,nascimento19}.

\begin{figure}
	\centering
	\hspace*{-0.5cm}\includegraphics[width=1\linewidth]{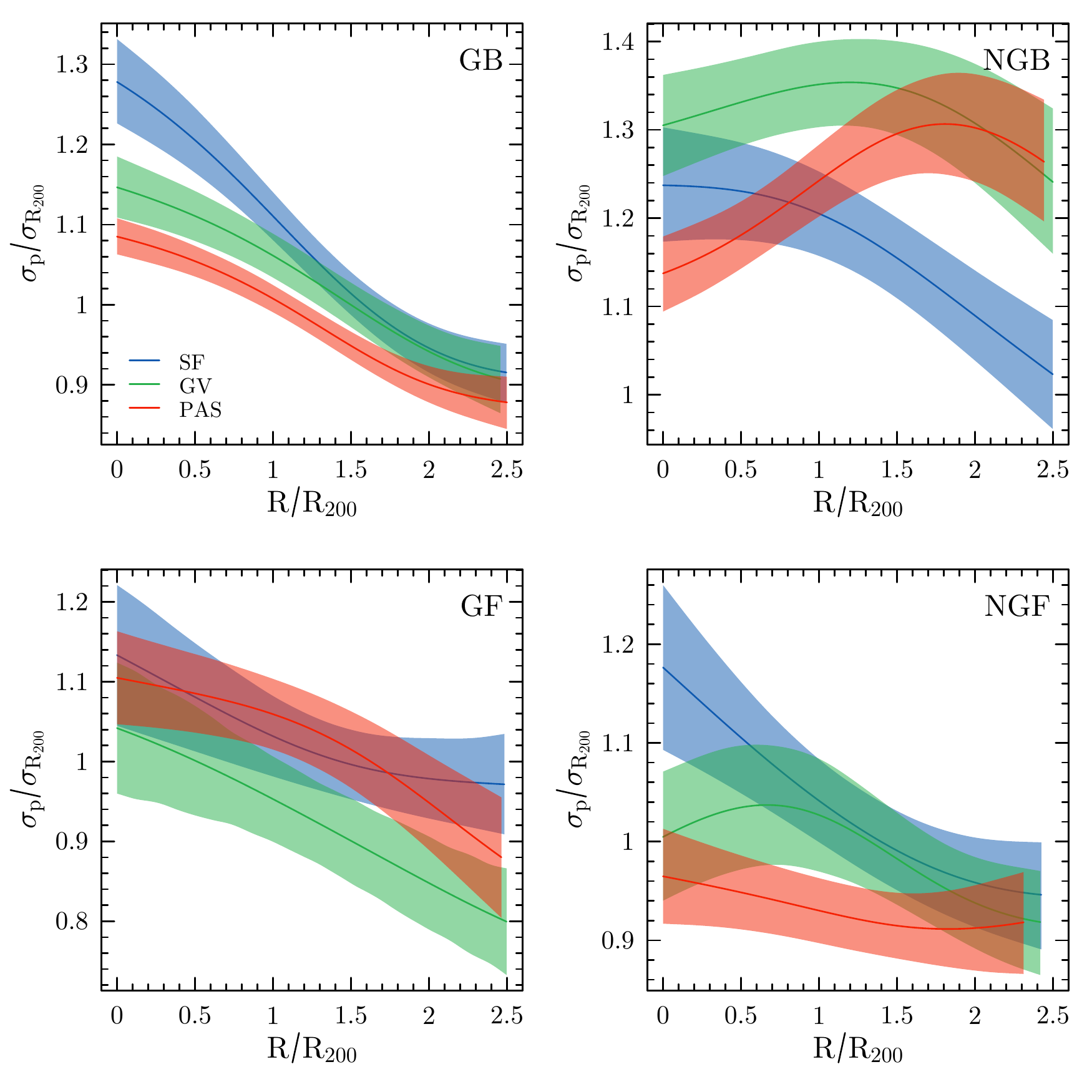}
 %  \vspace{-3cm}
	\caption{VDPs for our four subsamples, using the \citet{bergond06} prescription. Colors correspond to  evolutionary classifications. Shaded area represents the $ 1\sigma $ confidence interval.}
\label{fig8}
\end{figure}

%Once we have established evolutionary classes in previous sections, we therefore expect that their differences will also be reflected in their orbits. 

Galaxy classes and morphological changes should be related to galaxy orbits.
In this section we complement the analysis of the distinction between the evolutionary classes and morphologies from the dynamical point of view, comparing their radial velocity dispersion profiles (VDPs) in each sample as established before: GB, GF, NGB and NGF. The analysis of the VDPs is useful to indicate the degree of orbital segregation between the different evolutionary classes. 
The VDPs are constructed using the \citet{bergond06} definition, 
with the Gaussian window function corrected by \cite{bilton18}.
The window size was set to be the bi-weight scale estimator from \citet{beers90} for the values of the projected radii, thus overcoming the dependence on a somewhat arbitrary value, as used by other authors before \citep[see][]{bilton18}.
For each bin of $\rm R/R_{200}$ the dispersion and the corresponding $1\sigma$ confidence interval are determined from 1000 bootstrap realizations using the robust bi-weight estimators of location and scale from \citet{beers90}. Then, the VDPs were normalized by ${\rm R_{200}}$ and the velocity dispersion of the corresponding cluster.

%where the projected velocity dispersion is determined as:
%$$\sigma(R)=\sqrt{\frac{\sum_i\omega_i(R)(v_i-\bar{v})^2}{\sum_i\omega_i(R)}}$$
%with the Gaussian window function (corrected by \citeauthor{bilton18} \citeyear{bilton18}) with the negative exponential given by:
%$$ \omega_i(R)=\frac{1}{\sigma_R}\exp{\left[-\frac{(R-R_i)^2}{2\sigma_R^2}\right]} $$
%where $v_i$ and $R_i$ are the radial velocity and the radial clustercentric distance of the $i$-th galaxy, respectively, and $\bar{v}$ is the mean recession velocity of the sample. 
%We also determined the VDPs following the method described in \citet{costa17}, ordering galaxies by increasing radius and adding the corresponding velocities, one by one, starting from the first ten galaxies. At each radius, the projected velocity dispersion is determined using the bi-weight scale estimator for 1000 bootstrap realizations. Then, we estimated the mean and the one-sigma confidence interval, and also normalized the profiles in the same way as described above (Fig. 13). Note that this method is insensitive for tracing velocity dispersion at large radius, as galaxies in the outskirts modify very little the VDP shape; but on the other hand, it has no dependence on any arbitrary parameter. Altough the VDPs were obtained using different methods, the relationship between them within each subsample remains the same (see Fig. 12 and Fig. 13).

\begin{figure}
	\centering
	\hspace*{-0.5cm}\includegraphics[width=1\linewidth]{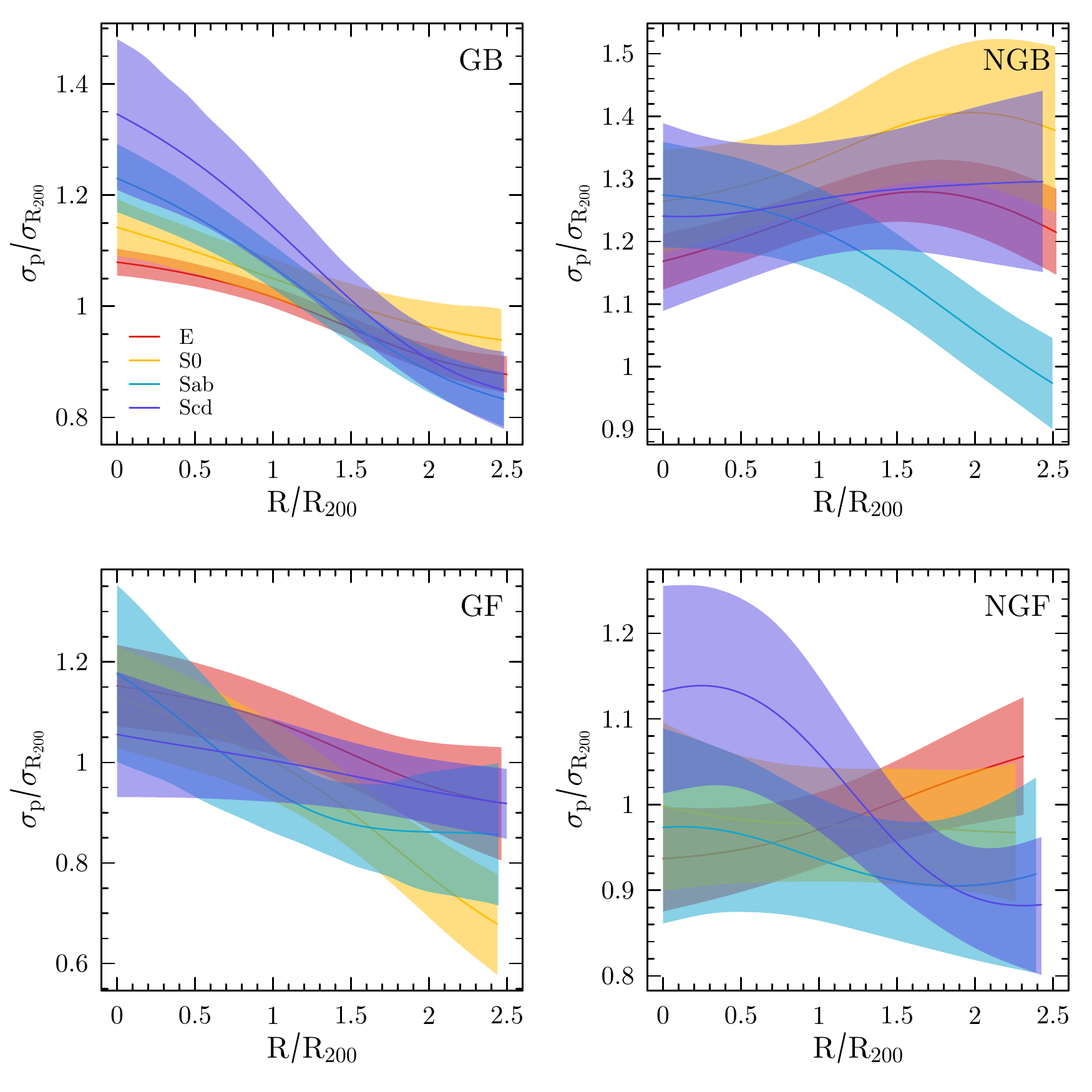}
%\vspace{-3cm}
	\caption{VDPs for our four subsamples, using the \citet{bergond06} prescription. Colors correspond to  morphological types. Shaded area represents the $ 1\sigma $ confidence interval.}
\label{fig9}
\end{figure}

\subsection{VDP results}

In Figure \ref{fig8} we observe that, for bright galaxies in G systems, VDPs for each evolutionary class are similar in shape, presenting a monotonic decreasing behaviour, typical of systems relaxed in the center plus a radial component from infalling objects \citep[e.g.][]{mohr1996optical, biviano2004eso,costa17}. Although the VDPs merge at large radii, they
are separated in the inner cluster regions, ${\rm \lesssim 1.5\;R_{200}}$, with SF, PAS and GV galaxies presenting 
higher, lower and intermediate velocity dispersions, respectively, which may be indicating different infall times for these populations.
Faint galaxies in G systems also present decreasing VDPs, but significantly entangled at all radii possibly because these galaxies have similar infall times with different levels of preprocessing, although the poorer statistics compared to the bright galaxies hampers a more sound conclusion.

A different picture can be seen for NG systems. For bright galaxies, the SF population still presents
a monotonic decreasing VDP, but both the GV and PAS populations now present increasing VDPs 
up to ${\rm \sim 1.5\;R_{200}}$, which may be indicating the presence of dynamic disturbances \citep{menci96,bilton18}.\footnote{Other possibility is circularization due to the two-body relaxation, but this is  efficient only in the very central regions of clusters
\citep[see e.g.][]{girardi1998optical}.}
As showed by \citet{fitchett1987substructure} interactions and substructures could modify the velocity dispersion profile in the central cluster region. In this case, a possible scenario is to assume that the time scale of the dynamic disturbance is shorter than the infall time of SF galaxies, which do not present the central inversion of the VDP.
In that context, note how passive galaxies is the most affected, with a marked inversion of the VDP, compared to that of the GV population. Still considering the VDP for  NGB systems, we should mention that the depression around ${\rm 1.0\;R_{200}}$ observed by \citet{costa17} possibly results from the superposition of the VDPs of SF, GV and PAS populations. Finally, faint galaxies in NG systems present the following composition: SF with decreasing VDP,
accompanied by a slightly declining (almost flat) VDP of PAS galaxies, and a disturbed VDP of GV galaxies.
This case is more complex, since we have an approximate flat component, usually associated with objects in completely isotropic orbits \citep[see][]{girardi1998optical}, combined with a disturbed GV. 

We can better understand the VDPs shown in Figure \ref{fig8} if they are reconstructed for the morphological types E, S0, Sab and Scd. This is presented in Figure \ref{fig9}, where we clearly  see, for GB systems, the gradation of profiles from that with the largest isotropic component (E galaxies) to the one with less isotropic orbit contribution (Scd galaxies). GF systems still present overlapping profiles that are not very enlightening for the discussion.
On the other hand, VDPs for NGB systems reveal that the component with  decreasing profile, previously associated to SF objects, corresponds to Sab galaxies, and this is the component that may be dominating the infall, after the others have experienced some dynamical interaction. At the same time, for  NGF systems, we see the only component with  decreasing profile
corresponding to Scd galaxies, while S0 and Sab galaxies present flat VDPs, and E galaxies have an inverted profile, possibly being the only disturbed component of the system. Since more evolved galaxies are also the ones that have been in the clusters for the longest time, we expect them to show signs of a dynamic disturbance before the other components, especially those that still enter the system in radial orbits. This could explain the undisturbed behavior of Sab and Scd galaxies in NGB and NGF systems, respectively.

%Comparing Figures \ref{fig8} and \ref{fig9}, we see that passive galaxies are not composed of E galaxies solely and that the GV of NGF systems should have some E galaxies.
%This could explain why GF systems present passive galaxies with star formation until $\sim$2 Gyr ago (as we have seen in Section 3.1).
% This result also recovers the idea that most evolved galaxies are also the ones that have been in the clusters for the longest time and, therefore, had a longer time to react to any type of dynamic disturbance.

\subsection{Asymmetry of infalling galaxies}

We must point out that bright galaxies infalling into NG systems are largely Sab, and that in the bright regime there is more conversion of Sab objects into S0, as we find in Section 3.4. In contrast, faint galaxies infalling into NG systems are typically Scd, and that in the faint regime there is more conversion of Scd objects into S0. This suggests a connection between galaxy interacions during the infall and the transformation of late type galaxies into S0. In the cluster environment several mechanisms (ram-pressure, harassment, starvation, etc.) can operate to start the transition of spirals to S0 \citep[see e.g.][and references therein]{boselli2006environmental,d2015transformation}. 
Disentangling these non-excluding mechanisms is beyond the discussion raised in this work, which is looking for differences between G and NG systems.

\begin{figure}
	\centering
	\hspace*{-0.5cm}\includegraphics[width=1\linewidth]{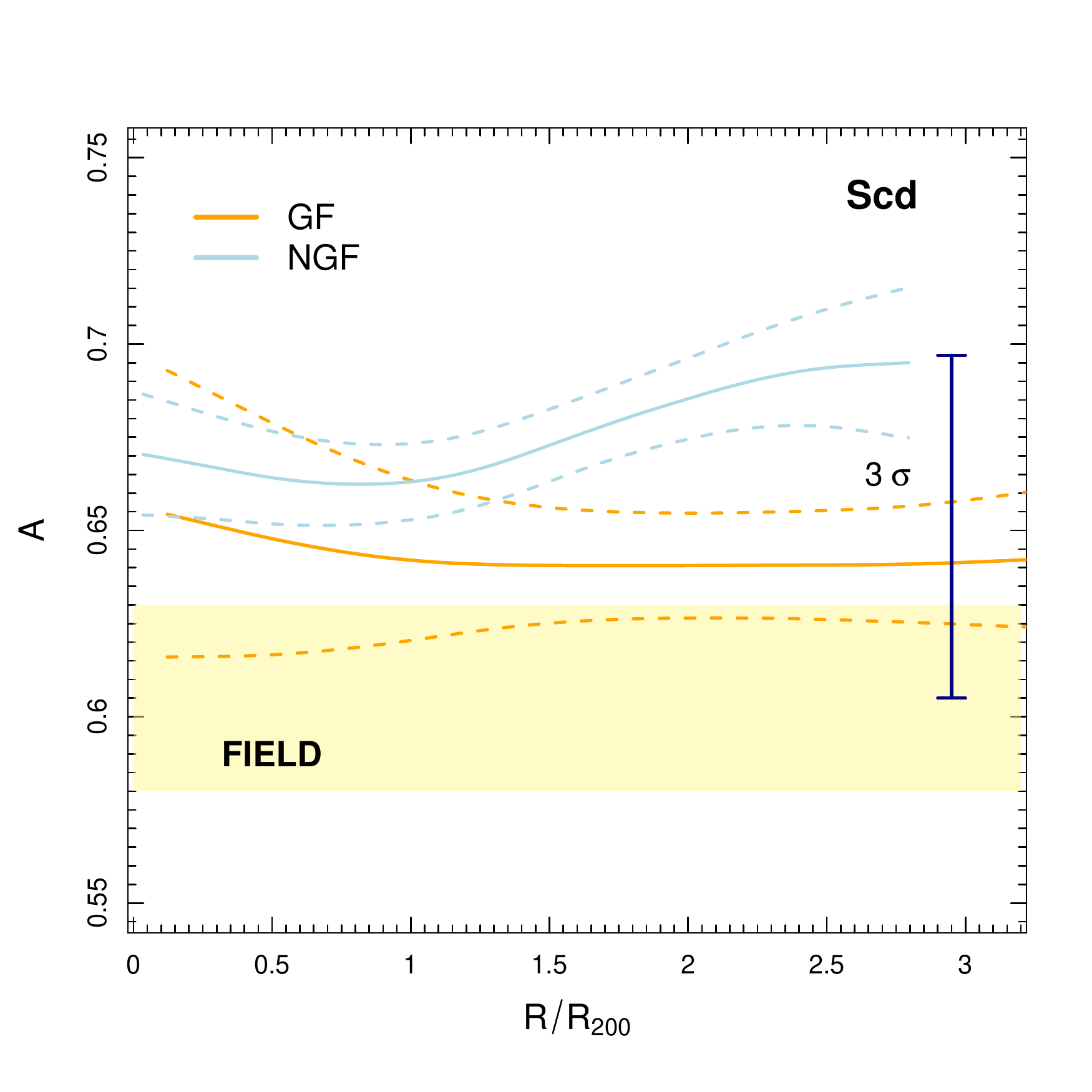}
\vspace{-0.5cm}
	\caption{Asymmetry parameter for faint Scd galaxies in G and NG systems.
Shaded areas indicate mean plus 1$\sigma$ asymmetries of Scd galaxies for a field sample in the same luminosity regime. Dashed lines indicate the $ 1\sigma $ confidence interval.}
\label{fig10}
\end{figure}

We use the asymmetry of galaxies as an indicator of interaction effects.
Asymmetry is measured using the correlation between the
original and rotated image with the code \textit{CyMorph} \citep[see][]{rosa18}.
Regarding bright Sab galaxies, we find no difference between G and NG systems $wrt$ bright
Sab in the field. For faint Scd we find an excess of asymmetry in NG systems. This is
shown in Figure \ref{fig10}, where the
shaded area indicates typical asymmetries of Scd galaxies for a field sample in the same luminosity regime as our faint samples.  
We should note a slight difference between faint Scd in G systems and Scd galaxies in the field (yellow shaded area). At the same time, faint Scd in NG systems present asymmetries varying from ${\rm \sim 2\sigma}$ at 1${\rm R_{200}}$ to  $\sim$3${\rm \sigma}$ at
3${\rm R_{200}}$ above the mean asymmetry in the field, suggesting they may be experiencing interactions that affect their shape. For low-mass galaxies, morphological transformation can be associated with minor mergers during the infall. In fact, \cite{moss2006enhanced} shows that 
$\sim 50-70$\% of the infall population are found to be in merging systems and slow galaxy-galaxy encounters. \cite{roberts2019quenching} propose a slow-then-rapid scenario of satellite quenching, with the low-quenching portion being consistent with quenching via steady gas depletion and the rapid-quenching  portion  being  consistent  with  ram-pressure stripping completing the quenching of low-mass satellites. Other possible mechanism is the harassment, which is characterized as the combined effect of tidal interactions between infalling galaxies and the global cluster  potential, and  of  the  tidal  interaction  with  close high-speed encounters with other cluster members \citep{moore1996galaxy}. 
%The efficiency of the harassment scenario is expected to depend on the orbit of infalling galaxies within a galaxy cluster \citep[e.g.][]{2015MNRAS.454.2502S}. The deeper an orbit is placed inside a cluster, the stronger the tidal force of the background mass of the cluster and the higher the probability of close encounters with other cluster members.

%possibly a combination
%of harassment and hydrodynamic galaxy-galaxy encounters.
%Since in Figure \ref{fig10} the behaviour of the asymmetry parameter is approximately constant within 
%0.5-2.0 ${\rm R_{200}}$,  we conclude that infalling Scd galaxies in NGF systems are not under the influence of a single mechanism, but of several gravitational and hydrodynamic processes.

\section{Summary}

In this work, we classify galaxies in the Age-SSFR space using the \textit{mclust} code. The procedure leads to three  galaxy populations: young with high SSFR (SF), old with low SSFR (PAS), and an intermediate population (GV).
Our aim is to find differences between these populations in the four samples of galaxies divided according to the dynamic state of their host clusters (G or NG) and the luminosity of the objects (B or F). Our main results are:

\begin{enumerate}
\setlength\itemsep{0.5em}
\item[(1)] Bright galaxies differ with respect to the populations identified in the Age-SSFR space. There are more bright SF $+$ GV galaxies in NG (55\% $\pm$ 4\%) than in G systems (43\% $\pm$ 3\%). We also find  more bright passive galaxies in G (57\% $\pm$ 2\%) than in NG (45\% $\pm$ 2\%) systems.
 Also, galaxies in GF and NGF samples present similar proportions of
types SF, GV, and PAS.

\item[(2)] The comparison of populations identified in the Age-SSFR space indicates strong difference between the populations (SF,GV,PAS), but weak or null difference between the samples G and NG, indicating that galaxies quench in the same way independently of the dynamical state of the clusters.

\item[(3)] Separating the intermediate population (GV) into star-forming (GV1) and passive components (GV2), we find more bright galaxies in the passive mode of NG (82\% $\pm$ 4\%) than in G (61\% $\pm$ 4\%) systems. 
We also find more intermediate faint galaxies in the star-forming side of NG (78\% $\pm$ 4\%) than in G (60\% $\pm$ 4\%) systems. This result indicates that the GV of G systems has a different composition compared to that of NG systems.

\item[(4)] Seven (of nine) parameters  exhibit, for all samples, the evolutionary sequence: ${\rm SF ~\rightarrow GV_{1} \rightarrow GV_{2} \rightarrow PAS}$, with significant differences of galaxies between these evolutionary stages. We also find a strong evolutionary segregation between bright and faint galaxies, independently on whether the galaxies are located in G or NG systems. This result indicates that the mechanisms of quenching  are more dependent on the luminosity (or mass) of galaxies, which agrees with several studies arguing in favor of 
stellar mass as the main driver of galaxy evolution \citep[e.g.][]{peng10,contini2019roles}.
Additional mechanisms are also required because mass-quenching alone cannot explain the morphological-structural transformations in the build up of passive galaxies or the various trends we have found.

\item[(5)] Fractions of galaxies E and S0 increase while those of Sab and Scd decrease along the
evolutionary sequence. We also find that GV1 is dominated by a mixture of S0 and Sab, while GV2 is more populated by S0 galaxies (and E galaxies in GB and NGB samples). This indicates that  morphological transformations of spiral galaxies into S0s must happen due to environmental effects,
both in G and NG systems.

\item[(6)] VDPs of bright galaxies in G and NG systems are quite different. In G systems
all populations present monotonic decreasing profiles, with differences only in the central
region of the clusters; while in NG systems PAS and GV populations seem to be dynamically
disturbed. Another important difference refers to the SF population in NG systems. Both bright and faint galaxies have decreasing VDP, indicating the presence of infall, but the bright ones correspond to Sab objects, while the faint are Scd galaxies. 

\item[(7)] Faint Scd in NG systems present asymmetries at least 3$\sigma$ above the mean asymmetry in the field, suggesting they may be experiencing interactions that affect their shape,
namely harassment, ram-pressure, and minor-mergers may be the mechanisms operating as the Scd galaxies feel the global cluster potential and start interacting with the cluster members. 

\item[(8)] Items (6) and (7) indicate that the main difference regarding the distinction of galaxy evolution in G and NG systems refers to the morphology of galaxies infalling onto these systems.

\end{enumerate}

\section*{Acknowledgements}
We thank the referee for useful suggestions.
We also thank R. Dupke for the critical reading of the work.
ALBR thanks the support of CNPq, grant 311932/2017-7.
RRdC acknowledges financial support from FAPESP through grant 2014/11156-4.
SBR thanks the support of FAPERGS.
PAAL thanks the support of CNPq, grant 309398/2018-5.
This study was financed in part by the Coordena\c c\~ao de Aperfei\c coamento de Pessoal de N\'{\i}vel Superior - Brasil (CAPES) - Finance Code 001.

This research has made use of the SAO/NASA Astrophysics
Data System, the NASA/IPAC Extragalactic Database (NED) and
the ESA Sky tool (sky.esa.int/). Funding for the SDSS and SDSS-II
was provided by the Alfred P. Sloan Foundation, the Participating Institutions, the National Science Foundation, the U.S. Department of Energy, the National Aeronautics and Space Administration, the Japanese Monbukagakusho, the Max Planck Society, and the Higher Education Funding Council for England.
A list of participating institutions can be obtained from the SDSS
Web Site http://www.sdss.org/.

\bibliographystyle{mnras}
\bibliography{citation}

%%%%%%%%%%%%%%%%%%%%%%%%%%%%%%%%%%%%%%%%%%%%%%%%%%

% Don't change these lines
\bsp	% typesetting comment
\label{lastpage}
\end{document}